\documentclass[preprint,aps, showpacs]{revtex4}
\usepackage{graphicx}
\usepackage{amsmath}
\usepackage[all]{xy}
\xyoption{arc}

\newcommand{\der}[1]{\frac{d #1}{d\lambda}}
\newcommand{\deru}[1]{\frac{d #1}{du}}
\newcommand{\derp}[1]{\frac{d^2 #1}{d\phi^2}}

\begin{document}

\title{Determining the Dimensionality of Spacetime by Gravitational Lensing}
\author{Jop Bri\"{e}t}
\email{j.briet@cwi.nl}
\affiliation{Centrum voor Wiskunde en Informatica, Kruislaan 413, NL-1098 SJ Amsterdam, The Netherlands} 
\author{David W. Hobill}
\email{hobill@crag.ucalgary.ca}
\affiliation{Department of Physics and Astronomy, University of Calgary, Calgary, Alberta, Canada, T2N 1N4}

\begin{abstract}
{The physics associated with spherically symmetric charged black holes
is analyzed from the
point of view of using weak gravitational lensing as a means for determining
the dimensionality of spacetime.  In particular, for exact solutions of electro-vac black holes
in four and five spacetime dimensions the motion of photons is studied
using the equations for the null geodesics and deriving the weak limit bending
angles and delays in photon arrival times.
}
\end{abstract}


\maketitle

\section{Introduction}

Recently there has been renewed interest in the higher order effects associated with gravitational lensing by
black holes in both the weak and strong field limits.
The motivations for such studies arise from both observational and theoretical
considerations \cite{will, eir, ser, bad, bcis,boz, boma, frit,vir}. Firstly, 
advances in the ability to perform high precision
measurements and observations are at the point where higher order effects are
close to being measurable. Secondly, many alternative gravity theories
(e.g.~those with higher dimensions, higher derivatives or higher powers of
curvature) differ from general relativity at second-order and beyond. In those
cases, the differences between the theories need to be understood in order to
guide what observations are to be made by experimental relativity practitioners. 

Within general relativity itself the properties of charge and rotation associated
with black holes only appear as higher 
order corrections to the Schwarzschild solution. Therefore in order to
determine what measurable effects such properties might have requires an
analysis that goes beyond simply utilising first order computations in a
perturbative analysis. 

The possibility that a black hole may be able to hold some non-zero electric
charge has been raised by a number of authors \cite{eir,ser}. In general relativity
the charge appears at second order in the vacuum Reissner-Nordstr\"{o}m solution
and therefore higher order computations are required in order to explore the 
differences between this and the 
uncharged Schwarzschild solution.  Charged black holes may well be
the end point of the evolution of massive, highly magnetized stars where the
neutralization of charge is avoided through some mechanism of selective
accretion \cite{puns}.  Isolated black holes may then be capable of remaining charged for
some time and may therefore be detectable through their influence on the passage
of light rays in the space surrounding them. Alternatively the
Reissner-Nordstr\"{o}m spacetime can act as an approximation to the
gravitational field of a slowly
rotating neutron star or other compact astrophysical object that has been
able to maintain a residual surface charge. Since
the Reissner-Nordstr\"{o}m solution is the unique spherically symmetric
electrovac solution in four spacetime dimensions, any gravitational lensing
effects that arise due to the presence of electrostatic charge from other
considerations should be compared against what one would expect in the 
Reissner-Nordstr\"{o}m spacetime.

However black holes (both charged and uncharged) appear in higher dimensional theories of gravity
as well. If spacetime is truly higher dimensional then black hole solutions in
those higher dimensions may make themselves known through the differences in the
expected physical behaviour of test particles in four dimensions. The question one might ask is 
whether it is possible to make
observations capable of measuring the differences between black holes
in four space time dimensions and those that might exist in
higher dimensions. 
For charged black holes in four-dimensions, these objects will 
have electromagnetic fields where the source of those fields are due to the existence 
of non-zero charges. However
higher dimensional black holes can have electromagnetic fields arising from
the purely geometric structure of the higher dimensions in the sense of the
original Kaluza-Klein theory.

In order to understand the physics that might arise as a result of 
gravitational collapse in higher dimensions we undertake a study of the gravitational
lensing of photons passing by charged black holes that are obtained as
electro-vacuum solutions in four dimensional Einstein theory and five-dimensional
(classical) Kaluza-Klein theory.  That is we compare the gravitational
lensing occurring in the vicinity of a Reissner-Nordstr\"{o}m black hole to that which
occurs due to a class
of charged black hole solutions in Kaluza-Klein theory that have
been discussed by Liu and Wesson \cite{liu}.  These higher dimensional black holes
can exist also without charge. However in that case the projection onto a
four-dimensional spacetime of the uncharged solution is equivalent to
the 4-dimensional Schwarzschild solution and no difference in gravitational
light ray bending would be measured. In the uncharged case the fifth dimension
is flat and will have no influence on the geodesic motion of test particles.
  Therefore in order to determine whether
or not the higher dimensional case might exist, it is necessary that the
black holes be capable of acting as a source of electric field and this 
will produce differences between the four and five-dimensional black holes solutions.
Since the fifth dimension is now non-flat it can be expected to have an influence on the motion
of photons in four dimensions.  This should lead to
differences in the light deflection angle and the ``Shapiro time delay''
suffered by photons passing by the black holes.

It has already been shown by Sereno \cite{ser} that the deflection angle of a
Reissner-Nordstr\"{o}m black hole is less than that for a Schwarzschild black hole
with the same mass.  That is the effect of the charge is to decrease the
deflection angle. This would result in lensed images that appear closer to the 
position of the lens and to each other.

A gravitational lensing observation alone is insufficient to determine
both the charge and the dimensionality of the black hole.  However should
the black hole have an accretion disk of ionized material surrounding it, one
can in principle determine the charge from the Lorentz force law. The electric
fields associated with the 4D and 5D charged black holes that are considered here
can be expected to take on their flat space Coulombic configuration at large distances and therefore 
the charge could be determined independently of the spacetime dimensions.

\section{Charged 5-D Kaluza-Klein black holes}

A number of spherically symmetric solutions to
the 5D Kaluza-Klein equations are known. Among the vacuum solutions many lack
event horizons and therefore cannot be considered as describing black hole solutions.
Alternatively, solutions with event horizons can be created by assuming a form of the metric that 
mimics the Schwarzschild solution but then these require a non-zero effective energy
momentum tensor on the right-hand-side of the field equations. Indeed a number of
authors \cite{majum, eir1, eir2} have studied gravitational lensing of
braneworld black holes where string tension is responsible for the formation of 
such objects. 
However, in what follows we will concentrate on a particular class of solutions that
have event horizons and are solutions to the 5D vacuum field equations.
In the appropriate limit such solutions reduce to the standard 4D Schwarzschild solution.
Some of the properties of these solutions
have been discussed previously by Liu and Wesson \cite{liu} who referred to
these objects as 5D charged black holes.

Using coordinates $(x^0, x^1,x^2,x^3,x^4)$ = $(t,r,\theta,\phi,\psi)$
where $\psi$ represents a spatial coordinate in the fifth dimension
together with the standard spherical polar (Schwarzschild or curvature) coordinates,
the line element for the charged black holes can be written in the form:
\begin{eqnarray}
ds^2 &=& \mathcal{B}(r)\mathcal{D}^{-1}(r) dt^2 - \mathcal{B}^{-1}(r) dr^2 - r^2 d\theta^2 - \nonumber\\
& &r^2\sin^2\theta d\phi^2 - \mathcal{D}(r) (d\psi+ \mathcal{A}(r) dt)^2 \label{equation:KKm}
\end{eqnarray}
where $\mathcal{A}(r)$, $\mathcal{B}(r)$, $\mathcal{D}(r)$ are ``potentials'' that can
be obtained by solving the five-dimensional, spherically symmetric, vacuum Einstein equations. 
As explicit functions of the radial coordinate, $r$, the potentials are also dependent upon two
arbitrary parameters $k$ and $M$. They may be written in the form (with
$G=c=1$):
\begin{eqnarray}
\mathcal{D} & \equiv & \frac{1-k\mathcal{B}}{1-k} = 1+ \frac{2Mk}{r} \\
\mathcal{A} & \equiv & \frac{\sqrt{k}(\mathcal{B}-1)}{1-k\mathcal{B}} = - \frac{2M\sqrt{k}}{\mathcal{D}r} \\
\mathcal{B} & \equiv & 1 - \frac{2M(1-k)}{r} = \mathcal{D}- \frac{2M}{r}
\end{eqnarray}
Of these potentials it can be shown that the potential $\mathcal{A}$ is the
solution to the Kaluza-Klein equations that are equivalent to the Maxwell
equations and this leads to a 
static radial electric field component which contributes to a Faraday tensor $F_{\alpha \beta}$ with a single
non-zero component:
$$ F_{01} = E(r) = \frac{2M\sqrt{k}}{\mathcal{D}^2 r^2}. $$
Using the expression for $\mathcal{D}$ and $\mathcal{A}$ and the condition that as $r \rightarrow
\infty$ the electrostatic potential must agree with the Coulomb potential one
has:
$$ \mathcal{A} \rightarrow -\frac{2M\sqrt{k}}{r} = -\frac{Q}{r}. $$
This in turn allows for the determination of the parameter $k$;
$$ k = \frac{Q^2}{4M^2}. $$

Therefore in terms of the charge $Q$ and the mass $M$ of the black hole, the
contributions to the metric coefficients are: 
\begin{eqnarray}
\frac{\mathcal{B}}{\mathcal{D}} & = & 1 - \frac{2M}{r + \frac{Q^2}{2M}} \label{equation:KKmc1}\\
\mathcal{B} & = & 1 - \big(2M - \frac{Q^2}{2M}\big)\frac{1}{r} \label{equation:KKmc2}\\
\mathcal{A} & = & -\frac{Q}{r+\frac{Q^2}{2M}} \label{equation:KKmc3} \\
\mathcal{D} & = & 1 + \frac{Q^2}{2Mr} \label{equation:KKmc4}
\end{eqnarray}
and the electric field becomes:
$$ E(r) = \frac{Q}{\mathcal{D}^2r^2}. $$
The projection of this metric onto the 4D subspace differs from the standard Reissner-Nordstr\"{o}m
solution even in the limit of large $r$  since the 4D line element of the
Kaluza-Klein solution has the approximate form:
\begin{eqnarray}
ds^2 & \simeq &
\left (1- \frac{2M}{r} + \frac{Q^2}{r^2} \right ) dt^2 -
\left(1- \frac{2M}{r} + \frac{Q^2}{2Mr} \right)^{-1} dr^2 \nonumber \\
 & &  - r^2 d\Omega ^2 \nonumber
\end{eqnarray}
where $d\Omega^2 = d\theta^2 + \sin^2 \theta d\phi^2$ is the 2-metric on the surface of the unit
sphere.
In practice, the electrostatic energy density will be small
($Gq^2/c^4r^2 \ll 1$) and the differences between the charged and uncharged
black hole solutions can be expected to be minor.
When $Q$ vanishes it is easy to see that the solution reduces to the
5D Schwarzschild vacuum solution which is just the 4D Schwarzschild
solution with a flat fifth dimension.

\section{Approximation of the deflection angle}

Since the contribution from the charge for the Reissner-Nordstr\"{o}m spacetime introduces
``second-order'' terms in the metric, the standard gravitational lensing and
time delay formulae must be extended to second order so as to include terms of
the form $M^2/r^2$, $Mr_0/r^2$, $Q^2/r^2$ etc.~where $r_0$ is a length scale that is of the
same order as $M$.   One can derive these either by expanding the
integrand that appears upon integrating the first order differential equation
form for the geodesic equations or by expanding the differential equations
themselves and solving these equations order by order.  The former technique
was 
employed by \cite{eir, ser, vir}, but we take the latter approach here in order
to verify the results obtained in those references.  Here 
it must mentioned that the higher order results may be coordinate dependent since the
deflection angle calculations are often computed with respect to the distance
of closest approach of the photon to the lens.  To first order the distance of
closest approach and the impact parameter are equivalent. However at second
order, they differ and the differences are coordinate dependent due to the fact
that unlike the impact parameter (which can be written in terms of the
constants of the motion and is therefore an invariant) the definition of the 
radius of closest approach will depend upon how one defines the radial
coordinate.

In this section the standard method of deriving a set of equations governing the motion of test
particles is used where we obtain a general second order differential equation for the 
inverse radial distance $u=1/r$ from the black hole as a
function of the azimuthal angle $\phi$. For the motion of null particles these equations has been 
derived by Bodenner and Will
\cite{will} for a Schwarzschild black hole 
using Schwarzschild, isotropic and harmonic radial coordinates. We follow this method
to derive the light bending angles associated with the Reissner-Nordstr\"{o}m,
and Kaluza-Klein black holes.  

One begins by recognizing that a general four dimensional, static, spherically symmetric line element
can be written as
\begin{equation}
ds^2 = A(r) dt^2 - B(r) dr^2 - C(r)r^2(d\theta^2 + \sin\theta d\phi^2).
\end{equation}
The equations of motion can be obtained either by varying the Lagrangian for the motion of zero-mass
particles or directly from the null geodesic equations.
To simplify the calculations, the variation in the polar angle can
be set to zero since we are dealing with spherical symmetry and motion in the plane defined
by $\theta = \pi/2$. The equations of motion for this case are
\begin{align}
&{\cal{E}} \equiv A\der{t} \label{equation:tee}\\
&J \equiv C r^2\der{\phi} \label{equation:phi}\\
&0 = \der{}\left(2B\der{r}\right) + A^\prime \left(\der{t}\right)^2 - B^\prime \left(\der{r}\right)^2  \nonumber\\
&\mbox{\hspace{18pt}} -(Cr^2)^\prime \left(\der{\phi}\right)^2, \label{equation:are}
\end{align}
where ${\cal{E}}$ and $J$ are the constants of the motion associated with the energy and angular momentum of the photon, $\lambda$
is an affine parameter along the geodesic, and a 
prime represents a derivative with respect
to $r$. Substituting equations (\ref{equation:tee}) and (\ref{equation:phi}) into (\ref{equation:are}), making the substitution $u = 1/r$ and re-writing
the equation such that $\phi$ is the independent variable, we obtain a second order differential equation for the
inverse radial distance from the black hole $u$,
\begin{equation}
\derp{u} + \left(\frac{C}{B}\right) u = - \frac{1}{2}u^2\deru{}\left(\frac{C}{B}\right)
+ \frac{{\cal{E}}^2}{2J^2}\deru{}\left(\frac{C^2}{AB}\right) \label{equation:ugen}.
\end{equation}
One now introduces the impact parameter $b \equiv J/\cal{E}$. Once the metric
coefficients are specified explicitly,
equation~(\ref{equation:ugen}) can be solved by successive approximations to
find the angle of deflection. While the form of the metric for both the Schwarzschild and 
Reissner-Nordstr\"{o}m solutions allows one to use the form (9) and the geodesic equation (13),
it will be seen that the 5D metric (1) also leads to the same equations of motion (13)
for photons which will allow us to use the same method of solution for all three 
spacetimes.

\begin{figure*}[t!]
\begin{center}
\includegraphics[width= 12cm]{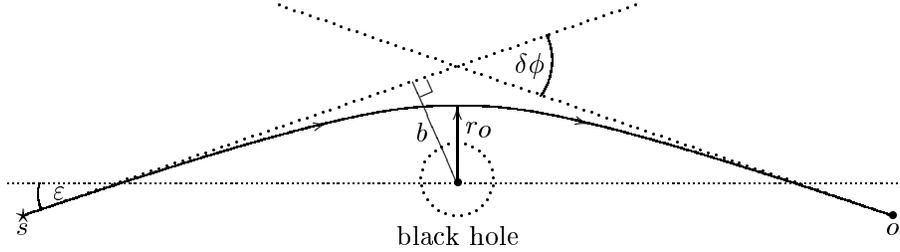}
\caption{\label{fig:defl} Photon deflection by spacetime curvature surrounding a black hole. The
angle $\phi =0$ when $r = r_0$. The impact parameter for the incoming light
beam is $b$.}
\end{center}
\end{figure*}

\subsection{Schwarzschild}
In Schwarzschild coordinates the metric coefficients are $A = B^{-1} = 1 - 2Mu$ and $C = 1$. This will leave equation~(\ref{equation:ugen})
as
\begin{equation}
\derp{u} + u = 3Mu^2 \label{equation:uSS}
\end{equation}
Assuming the solution is expandable in the form $u = u_0 + \epsilon u_1 + \epsilon^2 u_2 +\cdots$ will enable us to
approximate the solution to arbitrary order in $\epsilon$. The solution to the homogeneous (zeroth order) equation is
$u_0 = u_N\cos\phi$, where $u_N$ 
is equal to the inverse of the impact parameter ($u_N = 1/b$) which at this order is equal to the inverse of the distance of
closest approach. If we now set $\epsilon = Mu_N$, 
equation~(\ref{equation:uSS}) can be written as

\begin{align}
&(u_0'' + u_0) + \epsilon(u_1'' + u_1 - 3\cos^2\phi) +\nonumber\\
&\mbox{\hspace{18pt}} \epsilon^2(u_2'' + u_2 - 6 u_1\cos\phi) + \cdots = 0,
\end{align}
so that the equations up to second order in $\epsilon$ become

\begin{align}
&u_1'' + u_1 = 3 \cos^2\phi \\
&u_2'' + u_2 = 6 u_1\cos\phi
\end{align}

Solving these leaves us with the following expression for the inverse radial distance from the black hole,

\begin{align}
&u \simeq u_N\left(\cos\phi + \frac{1}{2}Mu_N[3 - \cos(2\phi)] +\right.\nonumber\\
&\mbox{\hspace{18pt}} \left.\frac{3}{16}M^2u_N^2[20\phi\sin\phi + \cos(3\phi)]\right).
\end{align}

When the light ray originates at a distant source located at $r_s$ and terminates at a distant observer located at
$r_o$, such that $r_{s,o}\rightarrow \infty$ or equivalently $u_{s,o}\rightarrow 0$ the deviation from straight line
motion is going to be very small. The deflection angle can be found by solving
for $\varepsilon$, using the source and observer angles $\phi_{s,o} = \pm(\pi/2 + \varepsilon)$ (See Fig.~\ref{fig:defl}). 
Since $\varepsilon \ll 1$, the trigonometric functions can be expanded in
powers of $\varepsilon$.  The trajectory is
symmetric about $\phi = 0$, so that the total deviation from straight line motion is the angle $\delta\phi = 2\varepsilon$. 
The expression for the deviation angle to second order in $\epsilon = Mu_N =
M/b$ can then be found to be:
\begin{equation}
\delta\phi \simeq \frac{4M}{b} + \frac{M^2}{b^2}\frac{15}{4}\pi.
\end{equation}
However the bending angle is often written in terms of the distance of closest
approach $r_0$ which is the value of $r$ when $\phi =0$.  This leads to the
relation:
$$1/r_0 = u_N + Mu_N^2 + 3M^2u_N^3/16$$
The approximation of the deflection angle up to second order $M/r_0$ now
becomes:
\begin{equation}
\delta\phi \simeq \frac{4M}{r_0} + \frac{M^2}{r_0^2}\left(\frac{15}{4}\pi - 4\right)
\end{equation}
It must be remembered that the second order terms in this latter expression is valid only for Schwarzschild 
coordinates since $r_0$ depends upon the choice of the radial coordinate
used to express the metric.  Since the impact parameter $b = J/{\cal{E}}$
depends only upon the constants of the motion, the expression (19) will be
invariant with respect to all choices of radial coordinates.  Of course the
first order expression which yields the ``Einstein angle'' is independent of
the radial coordinate since to this order $b \simeq r_0$.

\subsection{Reissner-Nordstr\"{o}m}

The metric coefficients for the Reissner-Nordstr\"{o}m solution are $A = B^{-1} = 1 - 2Mu + Q^2u^2$ and $C = 1$. Following the exact same
procedure as for the Schwarzschild case, equation (\ref{equation:ugen}) becomes
\begin{equation}
\derp{u} + u = 3Mu^2 - 2Q^2u^3\label{equation:uRN},
\end{equation}
so that the expanded form of equation (\ref{equation:uRN}) is
\begin{align}
&(u_0'' + u_0) + \epsilon(u_1'' + u_1 - 3\cos^2\phi) +\nonumber\\
&\mbox{\hspace{18pt}} \epsilon^2(u_2'' + u_2 - 6u_1\cos\phi + 2\frac{Q^2}{M^2}\cos^3\phi) + \cdots = 0.
\end{align}

Since $u_N$ is assumed to be small, the last term involving the charge $Q$ will be
second order in $\epsilon$ as long as $Q$ is less than or equal to $M$ (i.e.
there are no naked singularities present). 
The equations up to second order $\epsilon$ are now
\begin{align}
&u_1'' + u_1 = 3 \cos^2\phi \\
&u_2'' + u_2 = 6 u_1\cos\phi - 2\frac{Q^2}{M^2}\cos^3\phi,
\end{align}
so that the approximate solution for $u$ becomes:

\begin{align}
&u \simeq u_N\Bigg[\cos\phi + \frac{1}{2}Mu_N\Big(3 - \cos(2\phi)\Big) \:+\nonumber\\
&\frac{3}{16}M^2u_N^2\bigg(\Big(20 - \frac{4Q^2}{M^2}\Big)\phi\sin\phi + \Big(1 + \frac{Q^2}{3M^2}\Big)\cos(3\phi)\bigg)\Bigg],
\end{align}
and this yields a deflection angle expression in terms of the impact parameter:
\begin{equation}
\delta\phi \simeq \frac{4M}{b} + \frac{M^2}{b^2}\frac{15}{4}\pi - \frac{3}{4}\frac{Q^2}{b^2}\pi.
\end{equation}
or using the relation between the impact parameter and distance of closest
approach:
\begin{equation}
\delta\phi \simeq \frac{4M}{r_0} + \frac{M^2}{r_0^2}\left(\frac{15}{4}\pi - 4\right) - \frac{3}{4}\frac{Q^2}{r_0^2}\pi.
\end{equation}
The existence of the non-zero charge charge adds a small (but negative) correction in the second order term, which 
leads to a deflection angle that is smaller than that found in the Schwarzschild case.

\subsection{Kaluza-Klein}

The equations of geodesic motion for the Kaluza-Klein black hole spacetime will
be in terms of the 5-dimensional coordinates. 
Since the interest is
in the behaviour of photons in four dimensions, we need to determine
the equations in the 4D space.  The first step to solving the geodesic
equations is to determine what constants of the motion exist and this can be
most easily be accomplished by analyzing the Lagrangian associated with the
metric (\ref{equation:KKm})
\begin{align}
&\mathcal{L} = \frac{\mathcal{B}}{\mathcal{D}} \left( \frac{dt}{d\lambda}\right)^2 -
\frac{1}{\mathcal{B}}\left( \frac{dr}{d\lambda}\right)^2  \nonumber \\
& \mbox{\hspace{16pt}} - r^2\left[
\left(\frac{d\theta}{d\lambda}\right)^2 + \sin^2 \theta
\left(\frac{d\phi}{d\lambda}\right)^2 \right ] - \mathcal{D} \left[
\frac{d\psi}{d\lambda} + \mathcal{A} \frac{dt}{d \lambda} \right ]^2 
\label{equation:lagr},
\end{align}  
using the metric coefficients derived in section II. 
Here $\lambda$ is an affine parameter along the geodesic curve. As in the case
of both the Schwarzschild and Reissner-Nordstr\"{o}m solutions, the test
particle orbit remains in the plane defined by $\theta = \frac{\pi}{2}$. With $d\theta/d\lambda = 0$ 
the Lagrangian (\ref{equation:lagr}) leads directly to three constants of the motion:

\begin{eqnarray}
{\cal{E}} & \equiv & \frac{\mathcal{B}}{\mathcal{D}}\frac{dt}{d\lambda} -  \mathcal{D} \left[
\frac{d\psi}{d\lambda} + \mathcal{A} \frac{dt}{d \lambda} \right ]\mathcal{A} \\
J & \equiv & r^2 \frac{d\phi}{d\lambda} \\
N & \equiv &  \mathcal{D} \left[
\frac{d\psi}{d\lambda} + \mathcal{A}\frac{dt}{d \lambda} \right ]
\end{eqnarray}

The constants $J$ and $\mathcal{E}$ are, as before, related respectively to the angular
momentum and the energy of the photons whereas 
the constant of motion $N$ must be proportional to the charge $e$ of the test particle
in order to recover the Lorentz force law in the appropriate limiting
case (see \cite{liu}).

Since it is the motion of photons that are of interest here, the line
element and therefore the Lagrangian ${\cal{L}}$ must vanish. The test
particle charge also vanishes which leaves only two non trivial
constants of the motion along with the condition:
\begin{equation}
N = 0 \label{equation:KKmc5}
\end{equation}

Therefore the radial equation of motion becomes, after substituting in the
constants of the motion and the explicit expressions for the metric coefficients:
$$ \left(\frac{dr}{d\lambda}\right )^2  =
\left(1+\frac{Q^2}{2Mr}\right){\cal{E}}^2 - \frac{J^2}{r^2}\left[ 1 -
\frac{1}{r}
\left(2M-\frac{Q^2}{2M}\right) \right] .$$

This equation can be written in the form:
$$\left(\frac{dr}{d\lambda}\right ) ^2  + {\cal{V}}(r)  = {\cal{E}}^2 $$
where the `effective potential' takes the form:
$$ {\cal{V}}(r) = \left ( \frac{Q^2}{2M} - 2M \right) \frac{J^2}{r^3} +
\frac{J^2}{r^2} - \frac{Q^2{\cal{E}}^2}{2Mr} $$
which has a $1/r$ term for $Q \neq 0$. The effective potential
has a form similar to that for massive test particles in the Schwarzschild spacetime and therefore will lead to 
trapped photon orbits in the vicinity of the 5D black hole. That is, the
effective potential will have both a local minimum and a maximum.  
Unlike the 4D case,
one can have stable circular photon orbits for this particular Kaluza-Klein black hole.
Since the radius of the circular orbits are also energy dependent one would
expect to see a ``rainbow'' effect if one were to pass through the region of
stable photon orbits. 
As the charge approaches zero we just obtain the Schwarzschild configuration
where there is only an unstable orbit at a distance of three Schwarzschild
radii. For a non
zero charge we find two critical orbits; an unstable one closer to the black hole
and a stable one located further outward. These orbits can be found at
\begin{equation}
r_c = \frac{2MJ^2}{Q^2\mathcal{E}^2}\left[1 \pm \sqrt{1 - \frac{3Q^2\mathcal{E}^2}{J^2}\left(1 -
\frac{Q^2}{4M^2}\right)}\,\right]
\end{equation}          

Returning to the weak lensing case where $r$ is always well outside of the
the region close to the black hole we expect to obtain hyperbolic orbits
and we will now proceed to derive a scattering angle for such trajectories.

The equation for the
angle $\phi$ in terms of the radial position of the photon can be written as:
$$\frac{d\phi}{dr} = \frac{J/r^2}{\sqrt{ \mathcal{E}^2
+\displaystyle{\frac{Q^2\mathcal{E}^2}{2Mr}} -
\displaystyle{\frac{J^2}{r^2}} + \displaystyle{\frac{J^2}{r^2}}\left(2M -
\displaystyle{\frac{Q^2}{2M}}\right) }}$$
This result reduces to the 4D Schwarzschild result for photons when
$Q \rightarrow 0$.

With the metric coefficients given by (\ref{equation:KKmc1}), (\ref{equation:KKmc2}), (\ref{equation:KKmc3}) and
using (\ref{equation:KKmc5}), one obtains for the inverse radial coordinate:
\begin{eqnarray}
\derp{u} + u &=& \frac{Q^2}{4Mb^2} + \frac{3}{2}\left(2M - \frac{Q^2}{2M}\right)u^2\\ \label{equation:uKK}
	     &=& \alpha + \beta u^2.
\end{eqnarray}
Since we are going to approximate the solution to this equation only at distances larger than the impact
parameter, it can again be written in terms of a perturbation parameter $\epsilon$, such that $u'' + u = \alpha +
\epsilon(\beta u^2/\epsilon)$. The zeroth order solution to the now inhomogeneous equation is $u_0 = \alpha + u_N\cos\phi$.
If we now set $\epsilon = u_N\beta$, equation (\ref{equation:uKK}) becomes $u'' + u = \alpha + \epsilon(u_Nu^2)$.
Again expanding $u$ in terms of a power series in $\epsilon$, the equations to first and second order in
$\epsilon$ become:
\begin{align}
&u_1'' + u_1 = \frac{1}{b}\left(\alpha + u_N\cos\phi\right)^2\\
&u_2'' + u_2 = \frac{2}{b}\left(\alpha + u_N\cos\phi\right)u_1
\end{align}
respectively. After a straight forward but tedious calculation, the solution for the inverse radial distance can be written
explicitly in the form:
\begin{align}
&u \simeq \nonumber \\ 
&+ u_N\Bigg[\cos\phi + \frac{1}{2}Mu_N\Big(3 - \frac{Q^2}{4M^2} - \big(1 - \frac{Q^2}{4M^2}\big)\cos(2\phi)\Big) \nonumber\\
&+ \frac{3}{16}M^2u_N^2\Big(\big(20 - \frac{2Q^2}{M^2} + \frac{5Q^4}{4M^4}\big)\phi\sin\phi  \nonumber \\
&+ \big(1 - \frac{Q^2}{2M^2} + \frac{Q^4}{16M^4}\big)\cos(3\phi)\Big)\Bigg],
\end{align}
from which the deflection angle in terms of $r_0$ can be found to be
\begin{eqnarray}
\lefteqn{ \delta\phi \simeq \frac{4M}{r_0} - \frac{Q^2}{2Mr_0} + \frac{M^2}{r_0^2}\left(\frac{15}{4}\pi - 4\right) {} }\nonumber\\
& & +{}\frac{Q^2}{r_0^2}\left(1 - \frac{3}{8}\pi\right) + \frac{Q^4}{16M^2r_0^2}\left(\frac{15}{4}\pi - 3\right).
\end{eqnarray}
Or again using the impact parameter, one obtains
\begin{eqnarray}
\lefteqn{ \delta\phi \simeq \frac{4M}{b} - \frac{Q^2}{2Mb} + \frac{M^2}{b^2}\frac{15}{4}\pi {} }\nonumber\\
& & +{}\frac{Q^2}{b^2}\left(\frac{1}{2} - \frac{3}{8}\pi\right) + \frac{Q^4}{16M^2b^2}\left(\frac{15}{4}\pi - 3\right).
\end{eqnarray}

These expressions reduce to that obtained in the four-dimensional Schwarzschild
case when the charge vanishes.  This is to be expected since the fifth
dimension is then flat and will have no influence on the motion in the four
lower dimensions. However when the charge does not vanish, there is a
negative contribution to the deflection angle at first order and this should
have a significant effect on gravitational lensing as compared to the
Reissner-Nordstr\"{o}m spacetime.  The effect becomes even more significant for
small values of $r_0$ (or equivalently $b$) as will be shown in the next
section.

\section{Exact deflection angles}

\begin{center}
\begin{figure}[!]
\centering
\includegraphics[width=8.5cm]{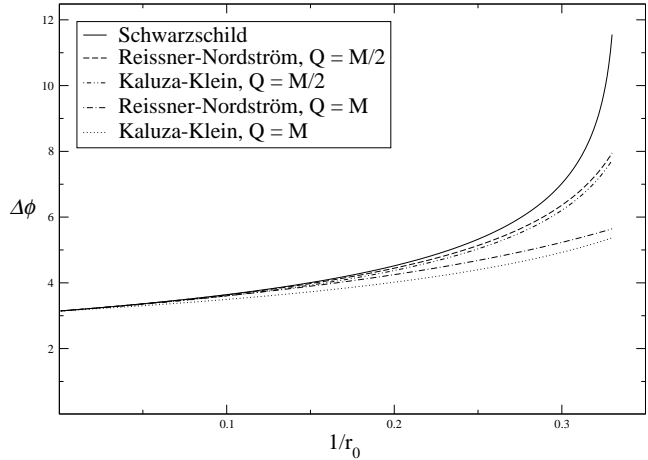}
\caption{\label{fig:angl} Total deflection angle $\Delta\phi$ (in radians) as a function of the inverse distance of closest
approach $1/r_0$ (in Schwarzschild radii).}
\end{figure}
\end{center}

The deflection angles can also be calculated exactly by finding an expression for the azimuthal angle in terms of
the radial distance. Following \S 8.5 of \cite{wei} we find that the total deflection angle can be found by
solving an integral in terms of the four dimensional metric coefficients.
\begin{equation}
\delta\phi = 2\int_{r_0}^{\infty} \frac{ \sqrt{ B(r)} }{ r\sqrt{ \left(\frac{r}{r_0}\right)^2 \left(\frac{A(r_0)}{A(r)}\right) - 1 }}dr - \pi
\end{equation}
Fortunately the metric coefficient associated with the fifth dimension does not come in into play for the motion of 
photons in the Kaluza Klein case since the last term in the
line element was found to be zero for uncharged test particles. One can evaluate the above integral numerically using 
a standard Simpson's rule method and setting the upper limit on the radial
position to $10^4 - 10^5$ times the value of $r_0$. A plot of the deflection angle $\Delta\phi = \pi + \delta\phi$
(See Fig.~\ref{fig:angl}) shows
that the deflection decreases as the charge on the black hole increases. This effect is clearly larger for the
five dimensional Kaluza-Klein solution than it is for the Reissner-Nordstr\"{o}m solution. When the charge on the black hole is zero, both the Reissner-Nordstr\"{o}m and
Kaluza-Klein solutions reduce to the Schwarzschild solution.

\section{Approximate time delay}
The deviations from the flat-space travel time (i.e. the Shapiro time delay) for photons passing by a black hole can
be estimated for all three geometries. Rather than take the approach that was used to compute the bending
angle, the alternative is to use the formal integral expression obtained from the
geodesic equation for the temporal coordinate, approximate the integrand to second order in the
appropriate expansion parameter and then evaluate the resulting integrals term
by term. 

Following \S 8.7 of \cite{wei}, we find that when curvature coordinates are used, the exact
time delay is given by
\begin{equation}
t{(r,r_0)} = \int^r_{r_0} \left(\frac{B(r)/A(r)}{1 -\left(\frac{A(r)}{A(r_0)}\right)\left(\frac{r_0}{r}\right)^2}\right)^{1/2}dr. \label{equation:intt} 
\end{equation}
Again some straightforward computations are required which are outlined in the
following subsections.

\subsection{Schwarzschild and Reissner-Nordstr\"{o}m spacetimes}

These spacetimes can be handled together by simply computing the results for the
Reissner-Nordstr\"{o}m solution and setting $Q=0$ to obtain the Schwarzschild
result.  

Expanding the metric functions in powers of $M/r$ and $Q/r$ leads to:
$$ A(r) = 1 - \frac{2M}{r} + \frac{Q^2}{r^2},  \;\;\;\; B(r) \simeq 1 + \frac{2M}{r} + \frac{\left(4M^2 - Q^2 \right)}{r^2}  $$
to second order.

The square of the denominator of the integral in equation (\ref{equation:intt}) takes the
form:
\begin{align}
& 1 - \frac{A(r)}{A(r_0)} \left(\frac{r_0}{r}\right)^2 \simeq \nonumber \\
&\mbox{\hspace{10pt}}\left(
1- \frac{r^2_0}{r^2} \right) \left[ 1 - \frac{2Mr_0}{r(r+ r_0)} -
\frac{4M^2}{r(r+ r_0)} + \frac{Q^2}{r^2} \right]  \label{equation:denom2}
\end{align}
so that equation(\ref{equation:intt}) expanded to second order becomes:
\begin{align}
&t{(r,r_0)} \simeq \int^r_{r_0} \left(1- \frac{r^2_0}{r^2} \right)^{-\frac{1}{2}} \times \nonumber \\
& \left[ 1 + \frac{2M}{r} +
\frac{Mr_0}{r(r+r_0)} + \frac{6M^2}{r^2} - \frac{3}{2}\frac{Q^2}{r^2} +
\frac{3}{2} \frac{M^2 r^2_0}{r^2(r+ r_0)^2} \right ] dr. 
\end{align}
Evaluating this integral term by term leads to:
\begin{align}
&t{(r,r_0)}  \simeq  \sqrt{r^2 - r^2_0} + 2M\ln \left( \frac{r+ \sqrt{r^2 - r^2_0}}{r_0} \right) \nonumber \\
& \mbox{\hspace{1pt}}+ M \sqrt{\frac{r-r_0}{r+r_0}} + \frac{3}{2}\frac{Q^2}{r_0}\left [\sin ^{-1}\left(\frac{r_0}{r}\right) 
- \frac{\pi}{2} \right ] \nonumber \\
& \mbox{\hspace{1pt}} +\frac{M^2}{r_0} \left[ \frac{15\pi}{4} - \frac{15}{2} \sin ^{-1}\left(\frac{r_0}{r}\right)
 - \frac{1}{2} \sqrt{\frac{r-r_0}{r+r_0}} \left(\frac{4r+5 r_0}{r+ r_0} \right) \right]. 
\end{align}

The first term represents the zeroth order (flat space) delay due to the photon
travel time from the distance of closest approach to the a position $r$ far
from the black hole.  The next two terms are the standard first
order ($M$ dependent) Shapiro time delays occurring in the Schwarzschild spacetime.  The second
order contributions to the time delay have a negative contribution from the charge
and additional positive terms from the mass.  In taking the limit where $r\gg r_0$ all the
second order effects introduce a constant delay 
$$\Delta t_2 = \frac{1}{4r_0}\left[ \left({15\pi} - 8 \right) M^2 - {3\pi}{Q^2}\right]. $$
Clearly these represent small deviations from
the standard Shapiro effect.

\subsection{Kaluza-Klein}

Since the time delay for photons moving in the 4D sector of the Kaluza-Klein spacetime
requires a knowledge of the metric coefficients $g_{00}$ and $g_{11}$ one
only needs to follow the methods presented in the preceding section.  In
particular we note that:
$$ A = 1 - \frac{2M}{r+\frac{Q^2}{2M}} \simeq 1 - \frac{2M}{r} + \frac{Q^2}{r^2} $$
which to second order agrees with the Reissner-Nordstr\"{o}m metric.
On the other hand we also have, upon defining $\kappa = 2M - Q^2/(2M)$
$$ B = \left[ 1 - \frac{\kappa}{r} \right]^{-1} \simeq  1 + \frac{\kappa}{r} +
\frac{\kappa^2}{r^2}. $$
Therefore we can use the expression (\ref{equation:denom2}) in the integral
(\ref{equation:intt}) and are only required to compute the ratio:
$$ \frac{B}{A} \simeq 1 + \frac{2M+\kappa}{r} + \frac{1}{r^2} [ \kappa^2 + 4M^2 - Q^2 + 2M\kappa] $$
which leads to an approximation for the integral
(\ref{equation:intt}):
\begin{align}
&t_{KK}{(r,r_0)}  \simeq  \int^r_{r_0} \left(
1- \frac{r^2_0}{r^2} \right)^{-\frac{1}{2}} \left[ 1 + \frac{1}{r} \left( 2M
-\frac{Q^2}{4M}\right) \right. \nonumber \\
 & \mbox{\hspace{10pt}}  + \frac{Mr_0}{r(r+r_0)} - \frac{Q^2r_0}{4r^2(r+r_0)}  \nonumber \\
& \mbox{\hspace{10pt}} \left. +  \frac{1}{r^2} \left(6M^2 - 2Q^2 + \frac{3}{32} \frac{Q^4}{M^2}\right)  
+ \frac{3}{2} \frac{M^2 r^2_0}{r^2(r+ r_0)^2} \right] dr. 
\end{align}
An evaluation of this integral gives the Kaluza-Klein time delay:
\begin{align}
&t_{KK}(r,r_0) \simeq \nonumber \\ 
&\sqrt{r^2 - r^2_0} +\left(2M - \frac{Q^2}{4M}\right)\ln \left( \frac{r+ \sqrt{r^2 - r^2_0}}{r_0} \right )
 \nonumber \\
& + M \left(
\frac{r-r_0}{r+r_0} \right)^{\frac{1}{2}} \nonumber \\
 & + \frac{M^2}{2r_0}\left[\frac{15\pi}{2} - 15
\sin^{-1}\left(\frac{r_0}{r}\right)
-\sqrt{\frac{r-r_0}{r+r_0}}\left(\frac{4r+5r_0}{r+r_0}\right)\right] \nonumber
\\
 & - \frac{Q^2}{4r_0}\left[\frac{9\pi}{2} - 9 \sin^{-1}\left(\frac{r_0}{r}\right)
+\sqrt{\frac{r-r_0}{r+r_0}} \right] \nonumber \\
 & + \frac{3}{32}\frac{Q^4}{M^2} \frac{1}{r_0}\left[\frac{\pi}{2} - \sin^{-1}\left(\frac{r_0}{r}\right) \right]
\end{align}

Again the zeroth order result is the flat space time delay as could be expected. The
the first order time delay has a negative contribution arising from the charge and
this significantly changes the time delay when compared with the result in the
Reissner-Nordstr\"{o}m spacetime, particularly in the case where the charge is
of the same order as the mass. Therefore the Kaluza-Klein charge
will produce a time delay that is significantly shorter than that which 
occurs in the Schwarzschild and Reissner-Nordstr\"{o}m cases. 
The second order contributions to the time delay 
due to the $M^2/r_0$ terms is the same as that found for the Schwarzschild
solution.  It is interesting to consider the full second order delay when
$r \gg r_0$: 
$${{\Delta t}_{KK}}_2 = \frac{1}{4r_0}\left[ \left({15\pi} - 8 \right) M^2 - \left(\frac{9}{2} 
- \frac{4}{\pi} - \frac{3}{16} \frac{Q^2}{M^2}\right ) \pi{Q^2}\right] $$
For the case $Q=M$ the numerical value of the factor in the curved brackets multiplying $\pi Q^2/4$ is 
$3.039\cdots$. This is very
close to the second order Reissner-Nordstr\"{o}m result of 3 given in eq.~(45).  

\begin{center}
\begin{figure}{!}
\centering
\includegraphics[width=8.5cm]{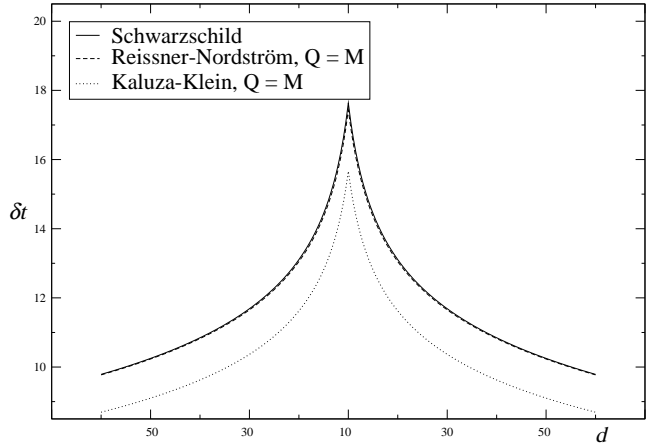}
\caption{\label{fig:delay} The time delay due to geometrical lensing as a function of the absolute value of the distance of closest approach 
of the light ray.
As a light source moves behind the black hole, the arrival time of its radiation increases. Here, $d$ 
measures the absolute value of the 
distance of closest approach ($d=|r_0|$) and ranges from 50M to 10M and back again.} \end{figure}
\end{center}

In Fig.~\ref{fig:delay} the time 
delays for the different
black holes are plotted as a function
of the absolute value of the distance of closest approach where the
distance between the source and observer is fixed at $10^5M$. For the charged
black holes $Q=M$. The units for
the time delay are in geometric units and will vary according to the distances
of the source and the observer
from the lens plane due to the logarithmic term appearing at first order.  This graph
clearly demonstrates that the second-order effects of the Reissner-Nordstr\"{o}m charge 
introduce only a small 
change to the overall time-delay compared to the Kaluza-Klein case where
the ratio $-Q^2/(2M)$ contributes to first-order.

\section{Discussion}

Without a Birkhoff type theorem for higher dimensional black holes,
the question of whether one might be able to determine the dimensionality of
spacetime through gravitational light bending and time delay observations will
clearly depend upon what one takes to be an appropriate black hole solution 
in higher dimensions. 
Many ``black hole'' solutions exist
in higher dimensional theories but these are often associated with non
vanishing effective energy-momentum tensors (for example in brane world
scenarios) that are capable of providing
solutions with black hole like characteristics in 4-dimensions.  The solution
studied here is a solution to the vacuum equations in 5-dimensions and in 
the spirit of the original Kaluza-Klein idea encodes the electromagnetic
vector potential into the higher dimensional metric tensor.

In this work one is simply asking the question of whether or not one
could in principle determine the difference between a 4-dimensional charged
black hole where the source of the electric field is a residual charge and
a 5-dimensional black hole where the electromagnetic potential seen in
4-dimensions is really an aspect of a higher-dimensional theory.

\begin{table*}
\caption{\label{tab:tab1} Deflection angles, $\delta \phi$, additional path lengths, $c\delta t$, and 
time delays, $\delta t$, for a photon trajectory starting and finishing at $10^5 M$  
from the black hole and with a distance of closest approach of $r_0= 10^2 M$. Here the mass units
are measured in terms of solar mass units where 
$M = Gm_\odot/c^2 \approx 1.4748$ km.}
\begin{center}
\begin{tabular}{|c|c|c|c|c|c|}
\hline \hline
& Schwarzschild & \multicolumn{2}{|c|}{Reissner-Nordstr\"{o}m} & \multicolumn{2}{|c|}{Kaluza-Klein} \\ \hline
& $Q=0$ & $Q= M/2$ & $Q= M$ & $Q= M/2$ & $Q= M$ \\ \hline
$\delta\phi$ (deg) & 2.33641267  & $\; $2.33303785$\; $ & $ $2.32291269$\; $ & $\; $2.26473475 $\; $ & $\; $2.05205798 $\; $ \\ \hline
$c\delta t$ (km) & 17.2313652 & 17.2227324  & 17.1968379 & 16.7431271 & 15.2800283 \\ \hline
$\delta t$ (msec) & 0.05743788  & 0.05740911 & 0.05732279 & 0.05581042 & 0.05093343 \\ \hline \hline
\end{tabular}
\end{center}
\end{table*}

It has been shown in this work that if the black hole is of the charged Liu-Wesson
type then one should be able to easily resolve the issue since the Kaluza-Klein 
black hole necessarily introduces
effects at first order in both the bending angle and the time delay where the charge 
can makes itself known to the observer.  While the charge might be determined
by an independent measurement, one could in practice attempt to determine 
the
bending angle, time delay and charge through measurements that would require a
best fit through the data.  This is what is usually done in gravitational
lensing
observations where the distance of closest approach or the impact parameter 
cannot be immediately determined without knowing the source position
and the initial direction of the photons themselves.

In order to show how large these effects might be for astrophysical black holes, Table~\ref{tab:tab1} is
included in order to provide a comparison of the light bending angles and the Shapiro delays
for black holes where $M$ equals one geometric solar mass unit. 
Geometric units are converted to the MKS system with the
appropriate factors of $G$ and $c$ re-introduced.  The distance of closest
approach of the photon trajectory is set at $r_0 = 100M$ and the time delay is
computed over a distance equal to $10^5 M$.  This is far enough from the black
hole so that no changes to the value of the bending angle occur to an accuracy of
one part in $10^7$.  The bending angles are given in degrees while the effective path
length that the photon must travel due to spacetime curvature and its
associated time delay are given in kilometres and milliseconds respectively.
As expected from the results shown in Figs.~\ref{fig:angl} and \ref{fig:delay}, even a significantly 
charged black hole of the Reissner-Nordstrom type makes only negligible
changes to the light bending angles and time delays.  On the other hand the charged
Kaluza-Klein black hole will produce very noticeable changes in the measurements
of such quantities.  

Of course, the time delay measurements require that the light source has some
known time dependent behaviour with a stable enough period that would allow one 
to compare the the differences in arrival time of signals from the source as it
passes behind the black hole.  Therefore one might take advantage of the
lensing of a variable star or (better yet) a pulsar \cite{hkp} in order to make
measurements that would test the dimensionality of space-time.

\begin{acknowledgements} 
The authors would like to thank L\'{a}szl\'{o} Gergely for
discussions concerning the nature of higher dimensional black holes. This
research was funded by an NSERC (Canada) Discovery Grant.
\end{acknowledgements} 


\end{document}